\documentstyle[12pt]{article}
\begin{document}

\baselineskip=22.pt plus 0.2pt minus 0.2pt
%\lineskiplimit=22.pt
\lineskip=22.pt plus 0.2pt minus 0.2pt
\def\half{\textstyle{1\over 2}}

\renewcommand{\thefootnote}{\fnsymbol{footnote}}

\setcounter{equation}{0}

\centerline{\Large\bf HBT Correlators --}
\centerline{\Large\bf Current Formalism vs.}
\centerline{\Large\bf Wigner Function Formulation\footnote[1]{Work
supported by BMFT, DFG, and GSI}}

\renewcommand{\thefootnote}{\alph{footnote}}

\vskip 0.8cm

\centerline{\large Scott Chapman and Ulrich Heinz}

\centerline{\em Institut f\"ur Theoretische Physik, Universit\"at
Regensburg,}
\centerline{\em D-93040 Regensburg, Germany}

\vskip 0.8cm
\noindent\underbar{\bf Abstract:}
We clarify the relationship between the current formalism developed
by Gyulassy, Kaufmann and Wilson and the Wigner function formulation
suggested by Pratt for the 2-particle correlator in Hanbury-Brown
Twiss interferometry. When applied to a hydrodynamical description of
the source with a sharp freeze-out hypersurface, our results remove a
slight error in the prescription given by Makhlin and Sinyukov which
has led to confusion in the literature.

\vskip .1 cm

\vspace*{.5in}

It is widely accepted that if the nuclear matter created in
ultra-relativistic heavy-ion collisions attains a high enough energy
density, it will undergo a phase transition into a quark-gluon plasma.
For this reason, it is of great interest to determine the energy
densities actually attained in these collisions. The total interaction
energy of a given reaction can be directly measured by particle
calorimeters and spectrometers. Although there is no analogous direct
measurement for the size of the reaction region, Hanbury-Brown Twiss
interferometry \cite{hbt} provides an indirect measurement in terms of
the correlations between produced particles.

Ten years ago, Pratt \cite{pratt} used the covariant current
formulation of Gyulassy, Kaufmann and Wilson \cite{gyul} to show that
the correlations between two particles could be expressed in terms of
one-particle pseudo-Wigner functions. Although Pratt's derivation was
non-relativistic, it provided a valuable link between the experimental
data and many semi-classical event generators whose output came in the
form of one-particle distributions. Since that time, different
methods have been used to relativistically generalize Pratt's result
\cite{csorgo,bertsch,padula}, but to our knowledge, the simplest
generalization (using the covariant current formalism covariantly) has
never been published. The aim of this letter is twofold: (1) to fill
the above void in the literature, and (2) to show that by applying the
final result to hydrodynamical models with 3-dimensional freezeout
hypersurfaces, a dispute in the literature about the correct form of
the 2-particle correlator in these models can be resolved.

The covariant single- and two-particle distributions for bosons are
defined by
 \begin{eqnarray}
   P_1({\bf p})
  & = & E\, \frac{dN}{d^3p}
        = E \, \langle\hat{a}^+({\bf p})\hat{a}({\bf p})\rangle \, ,
 \label{1} \\
   P_2({\bf p}_a,{\bf p}_b)
  & = & E_a\, E_b\, \frac{dN}{d^3p_ad^3p_b}
        = E_a \, E_b\, \langle\hat{a}^+({\bf p}_a)\hat{a}^+({\bf p}_b)
        \hat{a}({\bf p}_b)\hat{a}({\bf p}_a)\rangle\;,
 \label{2}
 \end{eqnarray}
where $\hat{a}^+({\bf p})$ ($\hat{a}({\bf p})$) creates (destroys) a
particle with momentum ${\bf p}$. The two particle correlation
function is then given by \cite{gyul}
 \begin{equation}
   C({\bf p}_a,{\bf p}_b)
   = \frac{\langle N\rangle^2}{\langle N(N-1)\rangle}\,
     \frac{P_2({\bf p}_a,{\bf p}_b)}{P_1({\bf p}_a)P_1({\bf p}_b)}\;.
 \label{3}
 \end{equation}

Using the classical covariant current formalism of \cite{gyul,kole}
we will show that for a general class of chaotic current ensembles
the two particle distribution for bosons obeys a Wick theorem:
 \begin{equation}
   P_2({\bf p}_a,{\bf p}_b)
   = \frac{\langle N(N-1)\rangle}{\langle N\rangle^2}\,
     \Bigl( P_1({\bf p}_a)P_1({\bf p}_b)
            + |\bar{S}({\bf p}_a,{\bf p}_b)|^2 \Bigr) \;,
 \label{3a}
 \end{equation}
where we define the following covariant quantity
 \begin{equation}
   \bar{S}({\bf p}_a,{\bf p}_b) = \sqrt{E_aE_b}\,
   \langle\hat{a}^+({\bf p}_a)\hat{a}({\bf p}_b)\rangle\;.
 \label{5}
 \end{equation}
We will then show that $\bar{S}$ is equal to the Fourier transform
of a kind of Wigner function:
 \begin{equation}
   \bar{S}({\bf p}_a,{\bf p}_b) =
   \tilde{S}({\bf q},{\bf K}) = \int d^4x\, e^{-iq{\cdot}x}\, S(x,K)\;,
 \label{6}
 \end{equation}
where the {\em off-shell} 4-vector $K=\half(p_a+p_b)$ is the average
of two on-shell ($p_i^0=E_i$) 4-momenta, and $q=p_a-p_b$ is the {\em
off-shell} difference of the same two momenta so that their scalar
product vanishes, $K^\mu q_\mu = 0$. For the special case of ${\bf
p}_a = {\bf p}_b$, $K=p_a$ becomes on-shell and
 \begin{equation}
   \bar{S}({\bf p}_a,{\bf p}_a)
   = \tilde{S}({\bf 0},{\bf K}) = P_1({\bf p}_a)
 \label{6a}
 \end{equation}
It should be noted that eqn.(\ref{6}) involves a 4-dimensional Wigner
transform, in contrast to the 3-dimensional expression suggested by
Pratt \cite{pratt} which neglects retardation effects.

In \cite{gyul} it was shown that a classical source current
$J(x)$ generates free outgoing pions in a state which satisfies
 \begin{equation}
   \hat{a}({\bf p})|J\rangle = i\tilde{J}({\bf p})|J\rangle \, ,
 \label{8a}
 \end{equation}
where
 \begin{equation}
   \tilde{J}({\bf p}) = \int \frac{d^4x}{\sqrt{(2\pi)^3 2 E_p}}\,
   \exp[i(E_pt-{\bf p}{\cdot}{\bf x})] \, J(x)
 \label{8}
 \end{equation}
is the on-shell Fourier transform of the source $J(x)$, and $\langle
J|J\rangle = \int d^3p\, \vert \tilde{J}({\bf p})\vert^2 = 1$. For
classical currents, the ensemble expectation values in eqns.
(\ref{1}), (\ref{2}), and (\ref{5}) can then be defined in terms of a
density operator $\hat{\rho}$ involving the state $|J\rangle$ such
that $\langle\hat{\cal O}\rangle={\rm tr}(\hat{\rho}\,\hat{\cal O})$.

Generalizing the result of \cite{kole} in order to allow for arbitrary
$x-p$ correlations, we consider an ensemble of chaotic source currents
at positions $x_i$ with momenta $p_i$,
 \begin{equation}
    J(x)
    = \sum_{i=1}^N e^{i\phi_i}\,e^{-ip_i{\cdot}(x-x_i)}\,
      J_0(x-x_i)\, ,
 \label{Jx}
 \end{equation}
where $\phi_i$ is a random phase. The momenta $p_i$ of the sources
can, but need not be on the boson mass-shell; for example, the source
could be a decaying $\Delta$-resonance with 3-momentum ${\bf p}_i$.
The on-shell Fourier transform of (\ref{Jx}) is
 \begin{equation}
    \tilde{J}({\bf p})
    = \sum_{i=1}^N e^{i\phi_i}\, e^{ip{\cdot}x_i}\,
      \tilde{J}_0(p-p_i)\, ,
 \label{8b}
 \end{equation}
where
 \begin{equation}
   \tilde{J}_0(p-p_i) = \int \frac{d^4x}{\sqrt{(2\pi)^3 2 E_p}}\,
   e^{i(p-p_i){\cdot}x}\, J_0(x)
 \label{J0}
 \end{equation}
is the Fourier transform of $J_0(x)$, and $p$ is on-shell while $p_i$
may be off-shell.

We then choose a density operator such that
 \begin{equation}
   {\rm tr}(\hat{\rho}\,\hat{\cal O})= \sum_{N=0}^\infty P_N
   \prod_{i=1}^N\int d^4x_i\, d^4p_i\,\rho(x_i,p_i)
   \int_0^{2\pi}\frac{d\phi_i}{2\pi}\,\langle J|\hat{\cal O}|J\rangle
 \label{8c}
 \end{equation}
where $\rho(x_i,p_i)$ is the covariant probability density of the
source points $(x_i,p_i)$ in phase space, and $P_N$ is the probability
distribution for the number of sources in the reaction. These
probabilities are normalized as follows:
 \begin{equation}
    \int d^4x\,d^4p\,\rho(x,p) = 1\, ,\qquad
    \sum_{N=0}^\infty P_N = 1 \, .
 \label{7b}
 \end{equation}

Using (\ref{8a}) and the above definitions, it is straightforward to
show that
 \begin{eqnarray}
   P_1({\bf p}) = E_p\,\langle|\tilde{J}({\bf p})|^2\rangle
   &=& \langle N\rangle E_p\int d^4x_1 \, d^4p_1\,
       \rho(x_1,p_1)\,|\tilde{J}_0(p-p_1)|^2
 \nonumber\\
   &=& \langle N\rangle E_p\int d^4p_1\,
       \tilde{\rho}(p_1)\,|\tilde{J}_0(p-p_1)|^2 \;.
 \label{8d}
 \end{eqnarray}
The single particle spectrum is thus obtained by folding the momentum
spectrum $\vert \tilde{J}_0(p)\vert^2$ of the individual source
currents $J_0$ with the 4-momentum distribution of the sources,
$\tilde{\rho}(p) = \int d^4x\, \rho(x,p)$.

Similarly, if one neglects cases in which two particles are emitted
from exactly the same point \cite{gyul}, one finds:
 \begin{equation}
    P_2({\bf p}_a,{\bf p}_b)
    = \frac{\langle N(N-1)\rangle}{\langle N\rangle^2}E_aE_b
      \biggl[ \langle| \tilde{J}({\bf p}_a)|^2\rangle
             \langle| \tilde{J}({\bf p}_b)|^2\rangle
          +  \langle \tilde{J}^*({\bf p}_a) \tilde{J}({\bf p}_b)\rangle
             \langle \tilde{J}^*({\bf p}_b) \tilde{J}({\bf p}_a)\rangle
      \biggr]
 \label{8d2}
 \end{equation}
which proves eqn.(\ref{3a}) by way of (\ref{8a}).

Using eqn.(\ref{8}), we find the following relationship:
 \begin{eqnarray}
   \tilde{J}^*({\bf p}_a)\, \tilde{J}({\bf p}_b)
   & = &
   \int \frac{d^4x_1\, d^4x_2}{(2\pi)^3 \,2\sqrt{E_aE_b}}\,
   \exp(-ip_a{\cdot}x_1 +ip_b{\cdot}x_2)J^*(x_1)J(x_2)
 \nonumber \\
   & = &
   \int\frac{d^4x\,d^4y}{(2\pi)^3\,2\sqrt{E_aE_b}}\,
   \exp(-iq{\cdot}x-iK{\cdot}y)J^*(x+\half y)J(x-\half y)  \, ,
 \label{8e}
 \end{eqnarray}
where $x=\half(x_1+x_2)$ and $y=x_1-x_2$. The above relation proves
eqn.(\ref{6}) as long as the following expression for the Wigner
function is used:
 \begin{equation}
   S(x,K) = \int\frac{d^4y}{2(2\pi)^3}\, e^{-iK{\cdot}y}
   \left\langle J^*(x+\half y)J(x-\half y)\right\rangle \, .
 \label{8f}
 \end{equation}
The average on the r.~h.~s. is defined in the sense of eqn.(\ref{8c})
and can be evaluated with the help of the definition (\ref{Jx}) to
yield
 \begin{equation}
   S(x,K) = \langle N \rangle \int d^4z\, d^4q \,\rho(x-z,q) \,
   S_0(z,K-q) \, ,
 \label{S}
 \end{equation}
where
 \begin{equation}
   S_0(x,p) = \int\frac{d^4y}{2(2\pi)^3}\, e^{-ip{\cdot}y}
              J_0^*(x+\half y)J_0(x-\half y) \,
 \label{S0}
 \end{equation}
is the Wigner function associated with an individual source $J_0$.
Thus the one- and two-particle spectra can be constructed from a
Wigner function which is obtained by folding the Wigner function for
an individual boson source $J_0$ with the Wigner distribution $\rho$
of the sources. Eqn.(\ref{S}) is useful for the calculation of
quantum statistical correlations from classical Monte Carlo event
generators for heavy-ion collisions: $\langle N \rangle \rho(x,p)$ can
be considered as the distribution of the classical phase-space
coordinates of the boson emitters (decaying resonances or 2-body
collision systems), and $S_0(x,p)$ as the Wigner function of the free
bosons emitted at these points. Replacing the former by a sum of
$\delta$-functions describing the space-time locations of the last
interactions and the boson momenta just afterwards, and the latter by
a product of two Gaussians with momentum spread $\Delta p$ and
coordinate spread $\Delta x$ such that $\Delta x\Delta p \geq
\hbar/2$, we recover the expressions derived in \cite{padula}.

Using eqns. (\ref{3}) to (\ref{6}), our final result is then
 \begin{equation}
   C({\bf p}_a,{\bf p}_b) = 1 + R({\bf q},{\bf K})
 \label{9}
 \end{equation}
where the ``correlator'' R is given by
 \begin{equation}
   R({\bf q},{\bf K}) = \frac{|\tilde{S}({\bf q},{\bf K})|^2}
   {\tilde{S}({\bf 0},{\bf p}_a)\,\tilde{S}({\bf 0},{\bf p}_b)}\;.
 \label{10}
 \end{equation}

Equation (\ref{6}) is the starting point for a practical evaluation of
the above correlator. It should be noted that due to the on-shell
condition of (\ref{8}), it is impossible to reconstruct $S(x,K)$ from
the correlator in a model independent way. Thus any ana\-lysis of data
on $R({\bf q},{\bf K})$ necessarily involves suitable model
assumptions for $S(x,K)$, in particular for the $x-K$ correlations in
the source distribution. In most practical applications one takes for
$S(x,K)$ a classical (on-shell) phase-space distribution. In
hydrodynamical models, for example, this phase-space distribution is
taken as a local equilibrium Bose-Einstein distribution localized on a
3-dimensional freeze-out hypersurface $\Sigma(x)$ which separates the
thermalized interior of an expanding fireball from the free-streaming
particles on its exterior \cite{marb}:
 \begin{equation}
   S_\alpha(x,K) = \frac{2s_\alpha+1}{(2\pi)^3}
   \int_\Sigma
   \frac{K^\mu d^3\!\sigma_\mu(x^\prime)\,\delta^{(4)}(x-x^\prime)}
   {\exp\{\beta(x^\prime)[K{\cdot}u(x^\prime)-\mu_\alpha(x^\prime)]\}
   - 1 } \;.
 \label{11}
 \end{equation}
Here $s_\alpha$ and $\mu_\alpha$ denote the spin and chemical
potential of the emitted particle species $\alpha$, while $u_\nu(x)$,
$\beta(x)$, and $d^3\sigma_\mu(x)$ denote the local hydrodynamic flow
velocity, inverse temperature, and normal-pointing freeze-out
hypersurface element. Inserting this equation into (\ref{6a}), one
obtains the Cooper-Frye formula \cite{coop}
 \begin{equation}
   \tilde{S}({\bf 0},{\bf p}) = P_1({\bf p}) =
   \int_\Sigma p^\mu d^3\!\sigma_\mu(x)\, f(x,p)
 \label{12}
 \end{equation}
where we define the distribution function (for clarity we drop the
index $\alpha$ for the particle species)
 \begin{equation}
   f(x,p) = \frac{2s+1}{(2\pi)^3}\frac{1}
   {\exp\{\beta(x)[p{\cdot}u(x)-\mu(x)]\} -1}\;.
 \label{13}
 \end{equation}

For the numerator of the correlator,
 \begin{equation}
   |\tilde{S}({\bf q},{\bf K})|^2
   = \int_\Sigma K^\mu d^3\!\sigma_\mu(x)\,  K^\nu d^3\!\sigma_\nu(y)\,
     f(x,K)\,f(y,K)\,\exp[iq{\cdot}(x-y)] \, ,
 \label{14}
 \end{equation}
we find an expression which is very similar to the one given in
\cite{sinu}. There, however, each of the two distribution
functions under the integral featured on-shell arguments $p_a$ and
$p_b$, respectively, instead of the common (off-shell) average
argument $K$ as in (\ref{14}).  This error in \cite{sinu} can be
traced back to an inaccurate transition from finite discrete volumes
along the freeze-out surface $\Sigma$ to the continuum limit
\cite{chap}. Taking over this inaccuracy produces (in particular for
very rapidly expanding sources) unphysical \cite{timm} oscillations of
the correlator around zero at large values of ${\bf q}$
\cite{sriv,mayer} which are inconsistent with the manifestly positive
definite nature of the correlator (\ref{10}).

The symmetric form (\ref{14}) (in contrast to the asymmetric one given
in \cite{sinu}) allows one to replace the exponential by the cosine
and to split the expression into two {\em real} 3-dimensional
integrals:
 \begin{equation}
   \bigl|\tilde{S}({\bf q},{\bf K})\bigr|^2
   =    \bigl(\tilde{S}_1({\bf q},{\bf K})\bigr)^2 +
        \bigl(\tilde{S}_2({\bf q},{\bf K})\bigr)^2 \, ,
 \label{15}
 \end{equation}
where
 \begin{equation}
   \tilde{S}_{1,2}({\bf q},{\bf K})
   = \int_\Sigma K^\mu d^3\!\sigma_\mu(x)\, f(x,K)\,
      \left\{
      \begin{array}{r}\cos(q{\cdot}x) \\ \sin(q{\cdot}x) \end{array}
      \right\}.
 \label{16}
 \end{equation}
This facilitates the numerical evaluation of the correlator.

\end{document}